\documentclass[conference]{IEEEtran}
\IEEEoverridecommandlockouts
\usepackage{cite}
\usepackage{amsmath,amssymb,amsfonts}
\usepackage{algorithmic}
\usepackage{graphicx}
\usepackage{textcomp}
\usepackage{xcolor}
\usepackage{color}
\usepackage{epstopdf}
\usepackage{fancyhdr}
\begin{document}
\bstctlcite{IEEEexample:BSTcontrol}
\title{Inter-satellite Quantum Key Distribution at Terahertz Frequencies\\
	\thanks{This work is funded by UNSW, Northrop Grumman Corporation, and the China Scholarship Council.}
}

\author{\IEEEauthorblockN{Ziqing Wang$^1$, Robert Malaney$^1$, and Jonathan Green$^{2}$}
	\IEEEauthorblockA{
		\textit{$^1$School of Electrical Engineering \& Telecommunications, The University of New South Wales, Sydney, NSW 2052, Australia.} \\
		\textit{$^{2}$Northrop Grumman Mission Systems, San Diego, California, USA}\\
		$^{1}$ziqing.wang1@student.unsw.edu.au, r.malaney@unsw.edu.au $^{2}$Jonathan.Green@ngc.com\\
	}
}
\maketitle
\thispagestyle{fancy}

\begin{abstract}
	Terahertz (THz)  communication is a topic of much research in the context of high-capacity next-generation wireless networks.  Quantum communication is also a topic of intensive research, most recently in the context of space-based deployments. In this work we explore the use of THz frequencies as a means to achieve quantum communication within a constellation of micro-satellites in Low-Earth-Orbit (LEO). Quantum communication between the  micro-satellite constellation and high-altitude terrestrial stations is also investigated. Our work demonstrates that THz quantum entanglement distribution and THz quantum key distribution are viable deployment options in the micro-satellite context. We discuss how such deployment opens up the possibility for simpler integration of global quantum and wireless networks. The possibility of using THz frequencies for quantum-radar applications in the context of LEO deployments is briefly discussed.
	
\end{abstract}

\section{Introduction}
Quantum Key Distribution (QKD)  offers security that is, in principle, guaranteed to be unconditional through its underlying dependence on quantum mechanics \cite{unconditional_security}.
In recent years, researchers have been investigating satellite-based QKD schemes so as to extend the achievable QKD range between terrestrial stations \cite{Sat_QKD_research_2002,Sat_QKD_research_1,Sat_QKD_research_2,Sat_QKD_research_3,Sat_QKD_research_4,Sat_QKD_research_5}; work that has partially come to experimental fruition \cite{China1200km,proj_Micius}.
Beyond the large ($\sim500$Kg) quantum satellite deployments of \cite{proj_Micius}  other works have suggested that low-cost micro-satellite (10-50Kg) technology could be a key enabler for global-scale quantum communications \cite{micro_satellite_50kg,nano_satellite}. The basic idea of the latter paradigm is to launch many low-cost micro-satellites that form a QKD network in space that can connect at any point to terrestrial transceivers. Direct connections between micro-satellites, avoid the unfavorable atmospheric turbulence and absorption that hinder terrestrial-satellite connections \cite{beam_wandering2012,beam_wandering2016}. Maximizing the former connections and minimizing the latter is therefore a useful design philosophy.

Parallel to these developments in space-based QKD, next-generation wireless networks based on 5th Generation (5G) standards, currently under development,  point to the  use of
THz frequencies (0.1-10THz\footnote{Although 10THz is considered by many as the upper limit of the THz band, our studies will consider frequencies up to 50THz.}) as a key enabler of short-range (sub 100m) high-capacity ($>$10GB)  links \cite{THz_wireless,THz_bands_survey}. Although such wireless networks will be principally based on microwave frequencies (0.3-300GHz\footnote{Beyond 30GHz is normally referred to as  millimeter frequencies by the telecommunication community. The `low-THz' frequency band and `high-millimeter' frequency bands overlap according to the same community.}), THz high-capacity links are viewed as extremely useful short-range solutions for 5G. Very recently, the notion of short-range quantum communications at THz frequencies has also been discussed \cite{THzQKD} - a notion that would allow for direct integration of quantum communication into emerging wireless networks without the need for complex optical-microwave interfaces (see \cite{bright_future_ralph,Neda_mmWave}).

In general, THz communications are limited in range due to the absorption caused by the high-water concentrations normally prevalent at ground level \cite{THz_Link_1_imaging_atmosphere,THz_wireless}. However,  it is noted that long-range terrestrial-to-satellite communications at THz frequencies is possible if the atmosphere is dry enough;  mountain-based observatories and Antarctica-based ground stations being well-known examples where such communications are feasible \cite{THz_mountain_top,THz_Antarctica,THz_Antarctica_Obs}. Of course, in Low-Earth-Orbit (LEO) THz communications are entirely feasible since water concentrations there can be considered negligible.

It is the purpose of this work to explore the use of THz frequencies as a means for quantum communications within the paradigm of micro-satellite networks connected to terrestrial stations placed in low-humidity environments. Relative to the pure-optical means normally touted for space-based quantum communications,  such a THz-based scenario may offer some deployment benefits with respect to future integrations with terrestrial wireless networks.

\section{System Model}\label{SysModel}
For low frequency (GHz-THz) quantum communications through air,  a trade-off between thermal noise and absorption is at play. GHz frequencies are less susceptible to absorption, but more susceptible to thermal noise, and vice versa for THz frequencies. At a fundamental level the noise enters any quantum information-transfer through photonic losses being replaced by thermal quantum states (in the optical regime at room temperatures the thermal states can be approximated by vacuum states). In fact, due to this effect the authors in \cite{THzQKD} have shown that the achievable distance of THz QKD is limited to 200m at a temperature of 300K. However, the much lower temperature in space can significantly extend the achievable distance of THz QKD. In fact, currently the normal operating temperature of many space-based optical components such as Charged Coupling Devices (CCD) are 173K \cite{CCD_173K_1,CCD_173K_2}.  Also, a recent work on thermal analysis for space-based quantum experiments has shown that carefully-designed radiation shielding can passively cool down the satellite-board optical bench to a temperature of 27K~\cite{shielding}. As we show below, such lower temperature environments can greatly assist us in improving QKD ranges.

We focus here on Gaussian continuous-variable (CV) quantum communications ~\cite{CW_Rev,Neda_Survey}. In CV technologies, information is modulated onto the two quadrature variables, $q$ and $p$, of the quantized electromagnetic field which spans an infinite-dimensional Hilbert space. Statistically, a single-mode Gaussian state can be fully characterized by the first order moments (mean) $\{\bar{q},\bar{p}\}$, and the second order moments (covariance) of its quadratures. An $N$-mode quantum system has a 2$N$-dimensional vector of  first moments $\{\bar{q}_1,\bar{p}_1,\bar{q}_2,\bar{p}_2,\ ...\ \bar{q}_N,\bar{p}_N\}$, and a 2$N$-dimensional covariance matrix (CM).

Let us use $\hat{\psi}$ to represent a single-mode thermal state with an average photon number of $\bar n$. $\hat{\psi}$ is a zero-mean Gaussian state with CM ${M_{0}} = V_0 \mathbf{I}$, where $V_0 = 2\bar n + 1$ is its quadrature variance, $\mathbf{I} = diag(1,1)$, and  $1$ is the vacuum shot noise unit (SNU). According to the blackbody radiation law, the average photon number for a thermal state  is given by
\begin{equation}\label{Eq_Blackbody}
\bar n = [{{\exp \left( {hf/{k}T_e} \right) - 1}}]^{-1},
\end{equation}
where $h$ is Planck's constant, $f$ is the frequency of the mode, $k$ is Boltzmann's constant, and $T_e$ is the temperature of the environment (or system). From~(\ref{Eq_Blackbody}), it is evident that the number of photons increases with temperature. In fact, when the temperature $T_e$ is 0K, the average photon number $\bar n$ will be zero, and $\hat{\psi}$ will become a pure vacuum state with zero mean and unity variance.

In this work we define four typical temperatures, namely 296K (room temperature), 173K (normal operational temperature of typical satellite-boarded optical components), 30K (the state-of-the-art temperature for a shielded, passively cooled satellite), and 3K (the thermal background temperature in deep space).

\subsection{Channel model}
Satellite-based optical quantum communications normally involves information exchange through an atmospheric channel. The atmospheric turbulence potentially causes scintillation, absorption and beam wandering effects, giving rise to severe fading and degradation of any quantum entanglement~\cite{beam_wandering2012,beam_wandering2016}.
However, in the scenario of inter-satellite communication there is negligible absorption, and the beam wandering effect can be ignored, thereby enabling us to approximate the channel as having a fixed attenuation due solely to diffraction effects.   In the rest of the paper, we will use this diffraction-only channel model to characterize the photonic loss mechanism for inter-satellite THz communications.

The diffraction-only channel model is a simple beam broadening model where we assume the beam center is perfectly aligned to the center of the receiver aperture. The photonic loss in this model is merely caused by the size of the diffracted beam at the receiver aperture~\cite{beam_wandering2012}. The transmissivity can be thus calculated as~\cite{laser_book}
\begin{equation}\label{Transmissivity}
T=1-\exp \left( -\frac{2{{r_a}^{2}}}{{{w}^{2}}\left( z \right)} \right),
\end{equation}
where $z$ is the propagation distance of the Gaussian beam, $r_a$ is the radius of receiver aperture, and $w\left( z \right)$ is the beam radius at distance $z$.

Under the Gaussian approximation, given the minimum of the beam radius (the beam-waist), we can calculate the beam radius at a certain distance  as
\begin{equation}
w\left( z \right)={{w}_{0}}\times {{\left[ 1+{{\left( \frac{\lambda z}{\pi w_{0}^{2}} \right)}^{2}} \right]}^{\frac{1}{2}}},
\end{equation}
where $\lambda$ is the wavelength of the beam, and $w_0$ is the radius of the beam-waist~\cite{laser_book}.
However, in micro-satellite settings, the  beam-waist and the receiver aperture are impacted by realistic hardware constraints.  In the work reported here we set both the beam-waist radius $w_0$ and the receiver-aperture radius $r_a$ to 10cm.\footnote{It has been previously reported that a successful satellite-to-ground Discrete Variable (DV) entanglement distribution experiment was carried out between two ground stations separated by 1200km. The satellite had two on-board telescopes for the two downlink channels, with mirror sizes of 18cm and 30cm.~\cite{China1200km}.}

\subsection{Gaussian entanglement}
At optical frequencies, the two-mode squeezed vacuum (TMSV) state is the ideal entanglement resource since its quadratures are maximally entangled given its average photon number~\cite{CW_Rev}. A TMSV state with squeezing $r,\,\,r \in \left[ {0,\infty } \right)$ is a Gaussian state with zero mean and the following CM
\begin{equation}\label{TMSVCM}%
{M_v} = \left( {\begin{array}{*{20}{c}}%
	{v\mathbf{I}}&{\sqrt {{v^2} - 1} \mathbf{Z}}\\%
	{\sqrt {{v^2} - 1} \mathbf{Z}}&{v\mathbf{I}}%
	\end{array}} \right),%
\end{equation}
where $v = \mathrm{cosh} (2r)$ is the quadrature variance of each mode, and where $\mathbf{Z} = diag(1, - 1)$. The squeezing in dB is given by $- 10 \log _ { 10 } ( \exp ( - 2 r ) )$.

However, the preparation noise caused by thermal fluctuations cannot be ignored at THz frequencies. Consequently, we adopt the two-mode squeezed thermal states as our entanglement resource. Such a  quantum state is  a zero-mean  Gaussian state with squeezing $r$, and CM~\cite{STS}
\begin{equation}\label{TMS-thermal}
\begin{array}{l}
{M_t} = \left( {\begin{array}{*{20}{c}}
	{\alpha \mathbf{I}}&{\gamma \mathbf{Z}}\\
	{\gamma \mathbf{Z}}&{\beta \mathbf{I}}
	\end{array}} \right),\,\,\rm where\\
\\
\alpha = 2{{\bar n}_1}{\cosh ^2}(r) + 2{{\bar n}_2}{\sinh ^2}(r) + \cosh (2r),\\
\\
\beta = 2{{\bar n}_1}{\sinh ^2}(r) + 2{{\bar n}_2}{\cosh ^2}(r) + \cosh (2r),\\
\\
\gamma = \left( {{{\bar n}_1} + {{\bar n}_2} + 1} \right)\sinh (2r),
\end{array}
\end{equation}
and where ${{\bar n}_1}$ and ${{\bar n}_2}$ are the average photon numbers of the two input thermal modes $1$ and $2$, respectively. ${{\bar n}_1}$ and ${{\bar n}_2}$ can be calculated using~(\ref{Eq_Blackbody}). It is clear that when ${{\bar n}_1}={{\bar n}_2}=0$, the two-mode squeezed thermal state reduces to a TMSV state.

We use the  logarithmic negativity to measure the degree of entanglement. The logarithmic negativity of a two-mode Gaussian state $\rho$ with a CM of $M=\left[ \begin{matrix}
A & C  \\ {{C}^{T}} & B  \\ \end{matrix} \right]$ is given by ${E_{LN}}(\rho) = \text{max}  [0,-{\log _2}\left( \nu_{-} \right)]$, where $\nu_{-}$ is the smallest symplectic eigenvalue of the partial transpose of $M$. $\nu_{-}$ is given by $\nu _{-} ^2 = \left( {\Delta  - \sqrt {{\Delta ^2} - 4\det M} } \right)/2$, where $\Delta = \det A + \det B - 2\det C$~\cite{CW_Rev}.

\subsection{Quantum key distribution}
\begin{figure}[!t]
	\begin{center}
		{\includegraphics[scale=0.3]{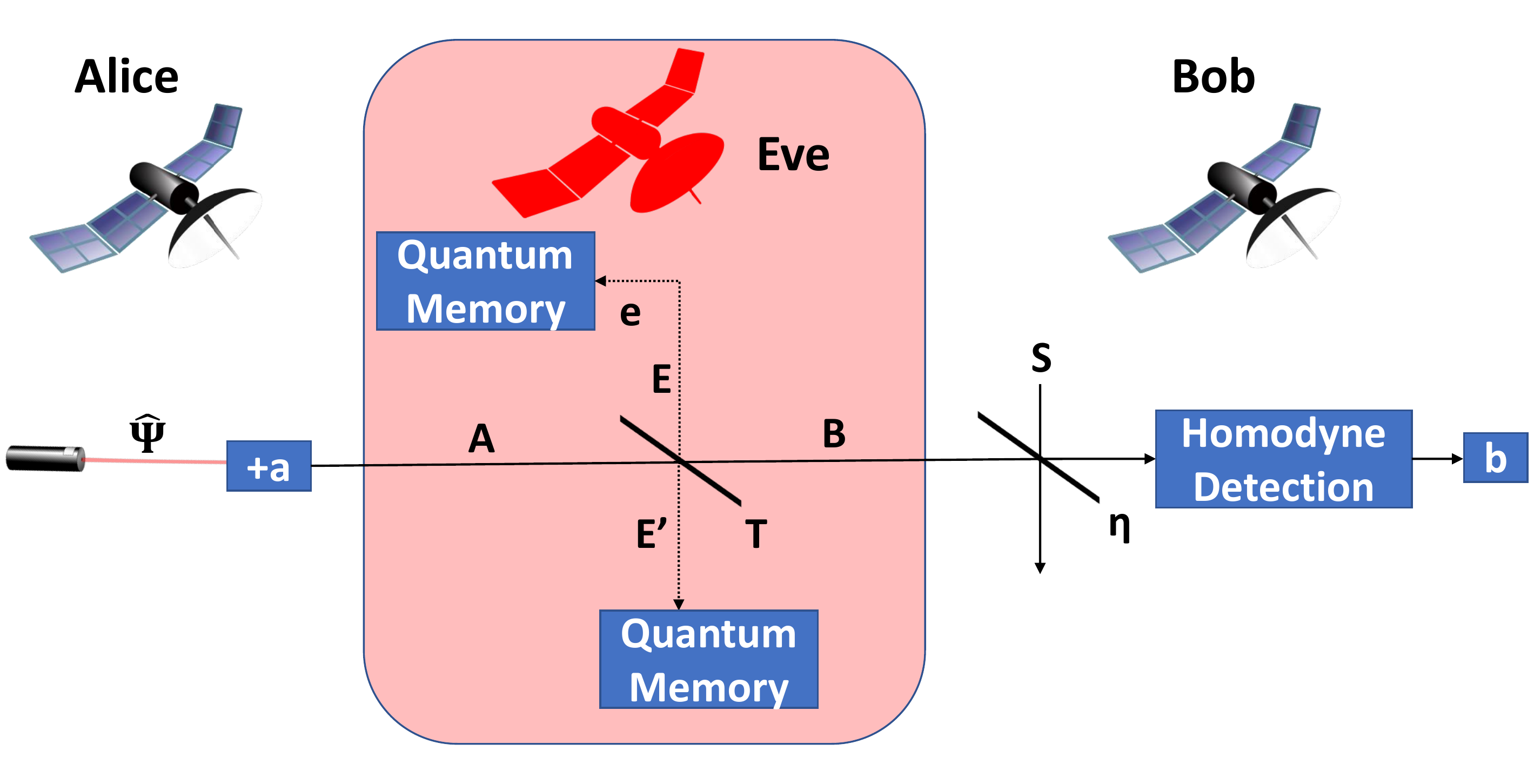}}
		\caption{The system model for inter-satellite THz QKD.}\label{QKD_Fig}
	\end{center}
\vspace{-0.6cm}
\end{figure}

For simplicity, we consider a prepare-and-measure (PM) scheme for QKD.
As is shown in Fig.~\ref{QKD_Fig}, let us assume that Alice prepares a Gaussian quantum state represented by the quadrature operator $A=\hat{\psi}+a$ with a total variance of  ${{V}_{A}}={{V}_{0}}+{{V}_{a}}$. Here $a$ represents the modulation (i.e. displacement) to a randomly-chosen quadrature ($q$ or $p$) of a thermal state $\hat{\psi}$ with variance ${V}_{0}$. The modulation $a$ is a zero-mean real Gaussian continuous variable with a variance of ${{V}_{a}}$ (in the following we will make the assumption ${{V}_{a}} \gg 1$).
Alice will then send all the modulated (displaced) states to Bob via a lossy inter-satellite THz channel. This channel can be modeled as a beam splitter with a transmissivity of $T$.
We assume that the channel transmissivity is calculated using~(\ref{Transmissivity}), and we also assume this channel is completely controlled by Eve.

In order to focus on the security problem, we assume Eve utilizes the collective entanglement cloner attack which is consistent with the most powerful attack against  Gaussian protocols in the asymptotic limit of infinite key length~\cite{in_the_middle,CW_Rev}. In such an attack, Eve prepares a TMSV state with variance $V_E$. Let us denote the two original modes in Eve's TMSV state as $\{e, E\}$. Then Eve's TMSV state has zero mean and a CM $V_{eE}$ in the form of~(\ref{TMSVCM}).

For every incoming state $A$, Eve uses one of the two modes, say $E$, to interact with Alice's transmitted quantum state $A$, creating two output modes $B$ and $E'$. Then, the output mode $B$ will be forwarded to Bob.  Eve's output mode $E'$, along with her original mode $e$, will be stored in her quantum memory. The pair of modes $\{e, E'\}$ will not be coherently detected by Eve until she has overheard all the classical information exchanged during the reconciliation process between Alice and Bob. We assume that Eve perfectly matches the preparation noise by setting $V_E=V_0$ so that Bob will perceive a total noise of $TV_0 + (1-T)V_E = V_0$. In this case, Alice and Bob cannot detect Eve. Due to Eve's entanglement cloner attack, the inter-satellite channel considered here becomes a thermal-loss channel~\cite{repeaterless_limit}.

During the measurement process, Bob uses a homodyne detector, and he randomly detects one quadrature of the incoming state $B$. Bob's realistic detector can also be modeled as a beam splitter with transmissivity $\eta$, which characterizes the detection efficiency. His detector is also subject to thermal fluctuations, and these fluctuations can also be modeled as a thermal state $S$ with a variance of $V_s$. In this paper we assume all micro-satellites share similar hardware configurations, therefore they are likely to be passively cooled down to the same temperature so that $V_s=V_0$. We denote Bob's measurement outcome as another random variable $b$.

After detection, a reconciliation process is necessary to make sure that Alice and Bob share the same information. There are two different reconciliation schemes, namely direct reconciliation (DR) and reverse reconciliation (RR). Previous work~\cite{THzQKD} has shown that the DR scheme can only support QKD at a very short distance. In this work we only consider the RR scheme. In the RR scheme, Bob reveals which quadrature variable he measured so that Alice can choose which realizations of $a$ should be kept.

\section{Simulation Results}

\subsection {Entanglement generation}\label{Section_Entanglement_Gen}
\begin{figure}[!t]
	\begin{center}
		{\includegraphics[width=3.3 in, height=2.5 in]{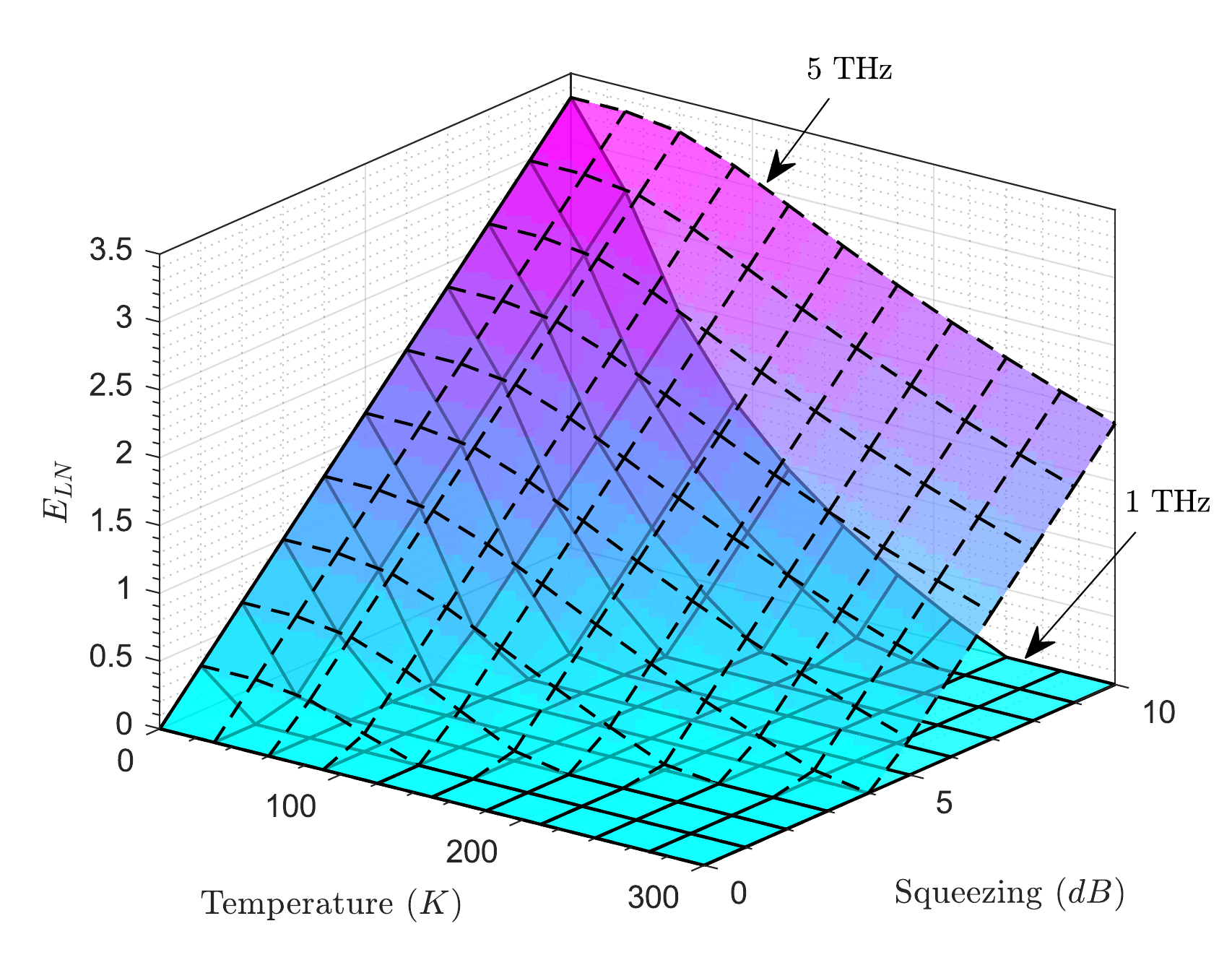}}
		\caption{The logarithmic negativity $E_{LN}$ of a two-mode squeezed thermal state against system temperature and squeezing for frequencies of 1THz and 5THz. It should be noted that no losses are assumed here.}\label{thermal_eff}
	\end{center}
\vspace{-0.5cm}
\end{figure}

We first study the influence that thermal fluctuations have on entanglement generation. To do this, we assume $\bar n _{1} = \bar n _{2}$. In Fig.~\ref{thermal_eff}, we consider two frequencies, 1THz and 5THz, and plot the logarithmic negativity $E_{LN}$ as a function of temperature and squeezing. As we can see from Fig.~\ref{thermal_eff}, it is clear the entanglement reduces with increasing temperature. However, this negative effect can be compensated by additional squeezing. We can observe that, for a squeezing of 10dB, the maximum temperature that allows for a non-zero entanglement is around 210K at 1THz. We also find that entanglement generation suffers less from thermal fluctuations at a higher frequency. In fact, we observe that, given enough squeezing (e.g. $r \ge 5dB$), it is even possible to generate entanglement at 5THz at 296K. However, the results here only provide the theoretical quality of the freshly-generated entanglement. After entanglement generation, signals are transferred. And during this process, realistic channel effects, such as attenuation will impose negative effects on the quality of entanglement as the signals propagate. The negative effects of a realistic channel will be discussed shortly.

For the reasons discussed earlier, we are mostly interested in two temperatures, 173K and 30K, under the state-of-the-art 10dB squeezing~\cite{10dB_Squeezing}. At 173K, we observe that a logarithmic negativity of 0.4 can be achieved at 1THz. We also observe that a logarithmic negativity of 2.5 can be achieved at 5THz. At 30K, these two values become 2.7 and 3.3 for 1THz and 5THz, respectively.

\subsection {Entanglement distribution}\label{Section_Entanglement_Dist}
We now study the performance of entanglement distribution at THz frequencies by investigating the evolution of the input entanglement resource $\rho_{in}$ in an inter-satellite channel. In the calculations of this sub-section we will ignore any blackbody radiation contributions from the channel. This means we will assume the transceivers are shielded perfectly from ambient backgrounds (e.g. the Earth's glow) and the 3K blackbody background is ignored.


Assuming the use of  state-of-the-art passive cooling, we set the temperature for entanglement generation in the satellite to 30K, and we set the temperature at all the satellite receivers to 30K as well. In this case, all communication parties considered here suffer from i.i.d. detection noise with the variance $V_s=2\bar{n}+1$, where $\bar{n}$ is calculated using~(\ref{Eq_Blackbody}) by setting $T_e=30 \rm K$. We also assume all the satellite-boarded receivers have the same detection efficiency of $\eta=10\%$.

We use the output entanglement $E_{LN}(\rho_{out})$ after the entanglement distribution process as the performance metric. When $E_{LN}(\rho_{out})=0$, we say that the channel is entanglement-breaking since no entanglement remains at the receiver output.


In our distribution scheme, Alice initially holds a two-mode squeezed thermal state. One mode (mode~1) is kept by her, and the other mode (mode~2) is transmitted to Bob via an inter-satellite THz channel. The resulting two-mode state is a zero-mean Gaussian state, and its CM can be written as
\begin{equation}\label{CM_single-mode}
{M_{AB}} = \left( {\begin{array}{*{20}{c}}
	{\alpha \mathbf{I}}&{\sqrt\eta\sqrt T  \gamma \mathbf{Z}}\\
	{\sqrt\eta\sqrt T  \gamma \mathbf{Z}}&{\beta'\mathbf{I}}
	\end{array}} \right),
\end{equation}
where $\beta'=\eta(T \beta + (1 - T ))+(1-\eta)V_s$, and $\alpha, \beta, \gamma$ are defined in~(\ref{TMS-thermal}). 

\begin{figure}[t]
	\begin{center}
		{\includegraphics[width=3.3 in, height=2.5 in] {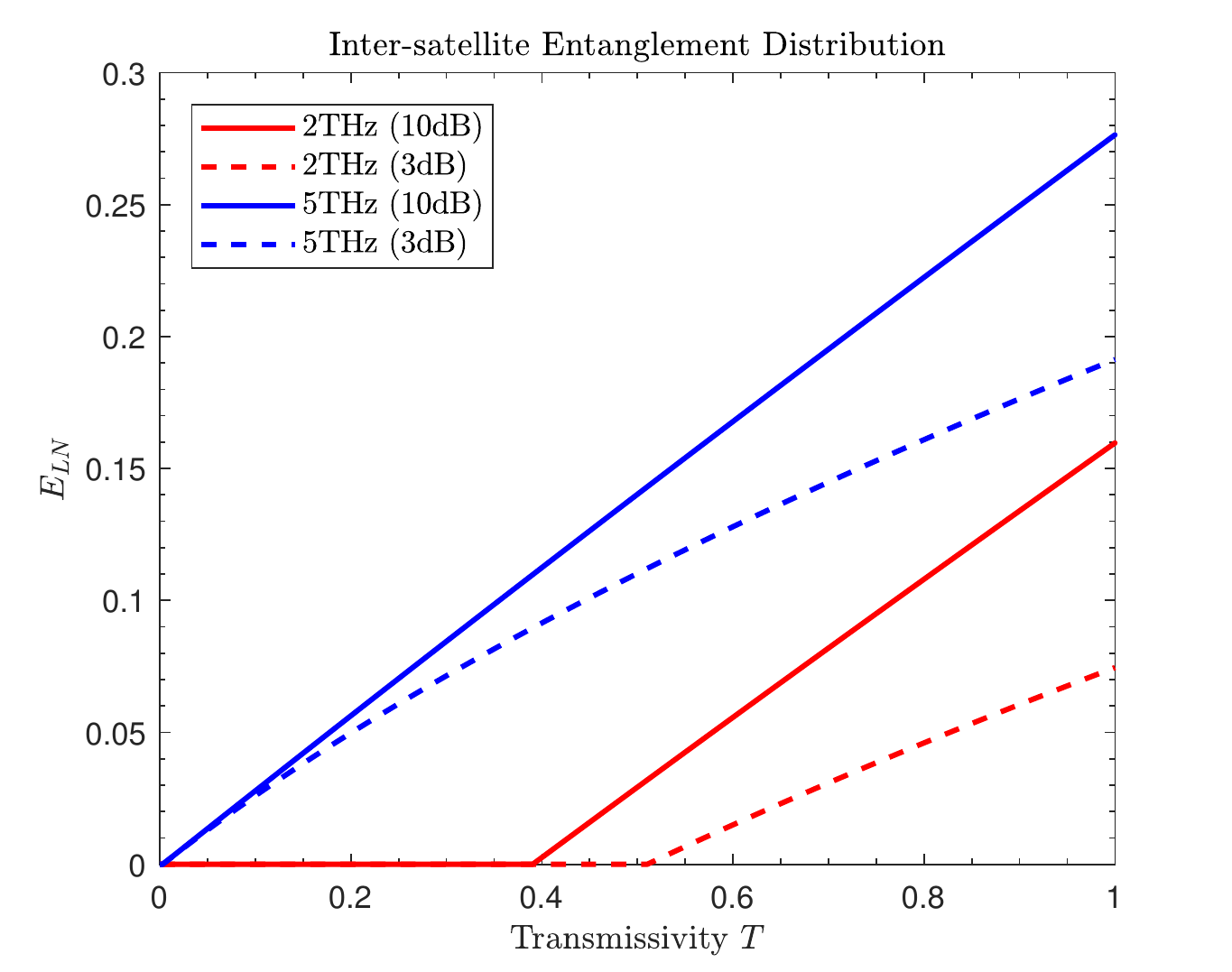}}
		\caption{The output entanglement $E_{LN}(\rho_{out})$ for inter-satellite entanglement distribution against the channel transmissivity $T$. The solid lines refer to 10dB initial squeezing, and the dashed lines refer to 3dB initial squeezing. Two THz frequencies, namely 2THz and 5THz, are considered.}\label{Fig_Entanglement_Distrubition}
	\end{center}
\vspace{-0.5cm}
\end{figure}

In Fig.~\ref{Fig_Entanglement_Distrubition}, we plot the output entanglement $E_{LN}(\rho_{out})$ for the two THz frequencies, namely 2THz and 5THz, against the channel transmissivity $T$. We also consider a 3dB and a 10dB initial squeezing at each frequency.

From Fig.~\ref{Fig_Entanglement_Distrubition} it is clear that the performance of entanglement distribution is reduced when the transmissivity decreases, and we can see that a higher frequency is more robust against the entanglement-breaking effect of the channel.   More specifically, we can see that the inter-satellite THz channel with a non-zero transmissivity retains a non-zero entanglement at 5THz frequency, even when a poor detection efficiency of $10\%$ is considered. However, it is not the same case at 2THz frequency, where a minimal transmissivity of 0.4 is needed for effective entanglement distribution for 10dB initial squeezing, and a minimal transmissivity of 0.5 is required for effective entanglement distribution for 3dB initial squeezing.

\subsection{Quantum key distribution}

 In this section we study the performance of inter-satellite QKD by studying the achievable secret key rate $R$. The secret key rate depends on how much information is leaked to Eve given the assumption that she overhears all classical communications between Alice and Bob. A positive key rate indicates that Alice and Bob manage to share more information than Eve~\cite{CV_thermal_state,in_the_middle,ApproachingClassicalLimits}.

\subsubsection{Secret key rate calculation}

Assuming optimal reconciliation, the RR-based secret key rate $R$ can be calculated as~\cite{KeyRate}
\begin{equation}%
R :=I(a:b)-I(E:b), \label{RR_Rate_t}
\end{equation}%
where $I(a:b)$ is the mutual information between Alice and Bob, and $I(E:b)$ is the information that Eve acquires after she has overheard the classical communications containing information related to $b$. Eve's information is bounded by the Holevo information $\chi(E:b)$
which is  is defined as~\cite{CW_Rev},
\begin{equation}
\chi:=H_{E}-H_{E|b}, \label{Holevo}%
\end{equation}
where $H_{E}$ and $H_{E|b}$ is the von Neumann entropy of Eve's total and conditional state (conditioned on the variable $b$), respectively. For Gaussian states, the von Neumann entropy is a simple function of symplectic eigenvalues~\cite{CW_Rev}, which reads
\begin{equation}%
H=\sum_{\nu}h(\nu), \label{h_vN_entropy}
\end{equation}
where $\nu \ge1$ are the symplectic eigenvalues of the CM of the Gaussian state, and
\begin{equation}
h(x):=\frac{x +1}{2}\log_{2}\frac{x +1}{2}-\frac{x -1}{2}\log_{2}\frac{x -1}{2}.
\label{h_vN_entropy_detail}
\end{equation}%

After some analysis (for details see~\cite{THzQKD}), we can express the key rate under the RR scheme using
\begin{align}
R(V_{0},T,\eta) & =h\left[  \sqrt{\frac{(1-T)\left(  1-\eta\right)  V_{0}+\eta}{\left(1-T\right)  \left(1-\eta\right)  +\eta}}\right]  -h( V_{0}) \nonumber\\
& ~+\frac{1}{2}\log_{2}\frac{(1-\eta)(1-T)+\eta}{1-T}.
\label{RR_Rate}%
\end{align}

We also compare the secret key rates with the fundamental upper bound on the channel capacity for an inter-satellite repeaterless thermal-loss channel with Gaussian thermal noise. This bound has recently been quoted as the Pirandola-Laurenza-Ottaviani-Banchi (PLOB) bound~\cite{repeaterless_limit,PLOB}. The PLOB bound is given by
\begin{equation}
\mathcal{C}=-\log_{2}[(1-T)T^{\bar{n}}]-h[V_{0}], \label{PLOB}%
\end{equation}
for $\bar{n}<T(1-T)^{-1}$, while $\mathcal{C}=0$ otherwise~\cite{repeaterless_limit}.

\begin{figure}[t]
	\begin{center}
		{\includegraphics[width=3.3 in, height=2.5 in]{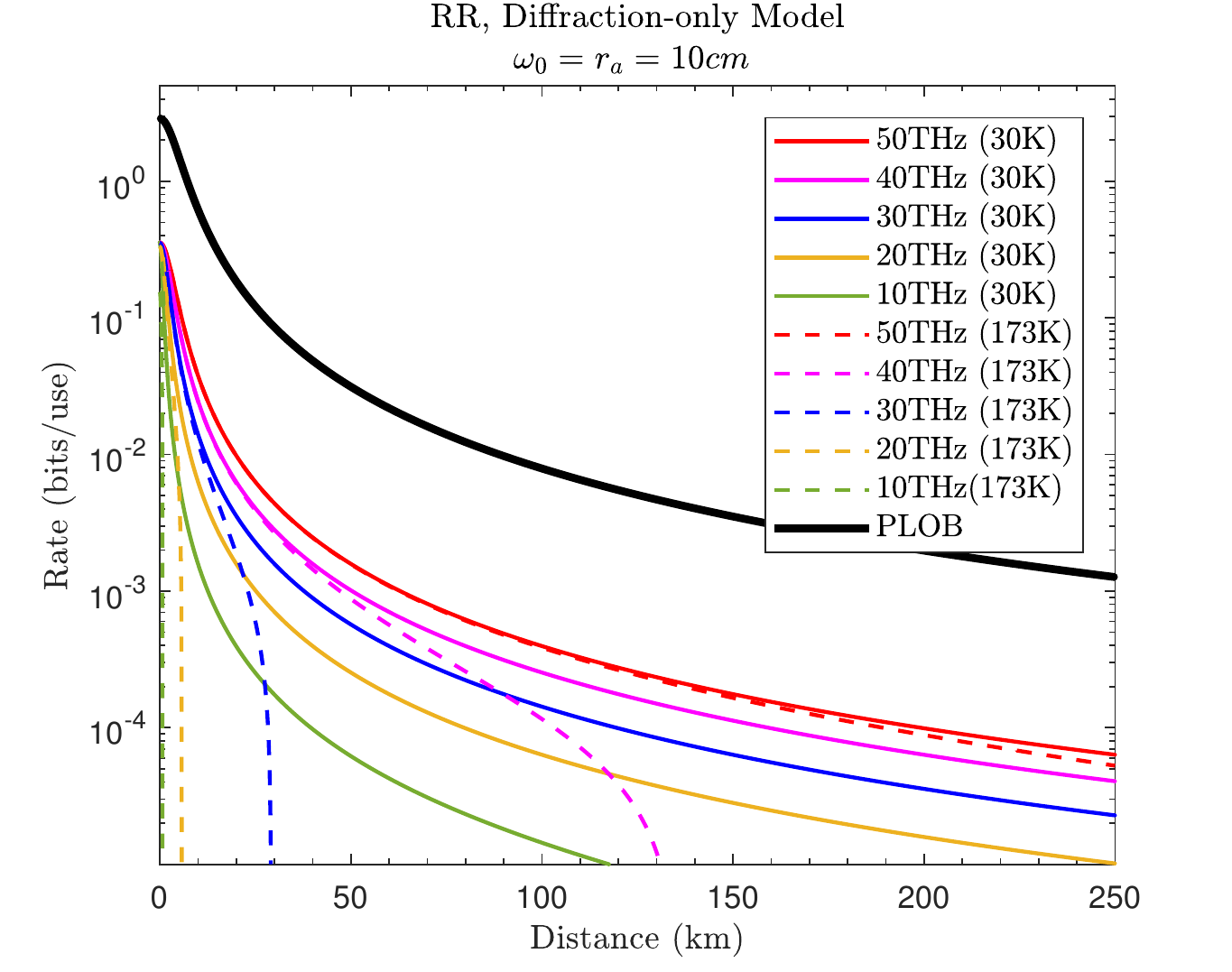}}
		\caption{The secret key rate of RR-based inter-satellite THz QKD protocol against distance for different THz frequencies. Here we set $\omega_0=r_a=10cm$. The solid lines refer to the temperature of 30K, and the dash lines refer to the temperature of 173K. The channel capacity bound (PLOB bound) is calculated using 50THz at 30K for comparison.}\label{RR_RateFig}
	\end{center}
\vspace{-0.0cm}
\end{figure}

We first set the detection efficiency to a reasonable 10\% in order to simulate a realistic hardware realization. In order to study the maximum achievable distance that the inter-satellite THz QKD can achieve, we plot the achievable secret key rate of RR-based inter-satellite THz QKD in bits/use (bits/channel use) against the distance for some typical THz frequencies in Fig.~\ref{RR_RateFig}. As we can see from this figure, the achievable secret key rate degrades exponentially when the QKD distance increases. This is due to the thermal noise  and the diffraction loss caused by the large beam size at the receiver. Also, we can see that the maximum achievable QKD distance is reduced when the temperature increases - a negative effect  significantly mitigated at higher frequencies.

Interestingly, at the state-of-the-art achievable operational temperature of 30K, we observe that a useful RR-based secret key rate of $10^{-3}$ bits/use can be achieved at tens of THz over some distance around 100km. We also observe that an RR-based secret key rate of $10^{-4}$ bits/use can be achieved over a 200-km-long inter-satellite 50THz channel, despite the low detection efficiency ($\eta = 10\%$). As such, the results shown  here indicate that a long QKD distance can be achieved over the inter-satellite channel. That is, the results strongly support the feasibility of inter-satellite THz QKD.

\subsubsection{Accessible frequency}
According to the blackbody radiation law in~(\ref{Eq_Blackbody}), the degree of thermal fluctuations is reduced when the frequency increases, and therefore a higher frequency is always preferred for QKD. But \emph{what is the minimum frequency that can be used to carry out effective inter-satellite QKD under a fixed temperature}? To answer this question, the accessible frequency is studied in this section. The accessible frequency here refers to the minimum frequency that can support a positive quantum key rate over an inter-satellite channel.
For the RR-based QKD protocol considered in this work, the accessible frequency can be found numerically. Since the channel capacity $\mathcal{C}$ has an upper bound, the accessible frequency has a corresponding lower bound. The lower bound of accessible frequency is achieved when the channel capacity reaches zero ($\mathcal{C}=0$), and this means $\bar{n}=T(1-T)^{-1}$~\cite{THzQKD}.
According to~(\ref{Eq_Blackbody}), this lower bound of accessible frequency can be given in the closed form of
\begin{equation}
f_{\min}=\frac{\ln(\frac{1}{T})  k  T_e}{h}. \label{AccessFreq_Bound}
\end{equation}

\begin{figure}[!t]
	\begin{center}
		{\includegraphics[width=3.3 in, height=2.5 in]{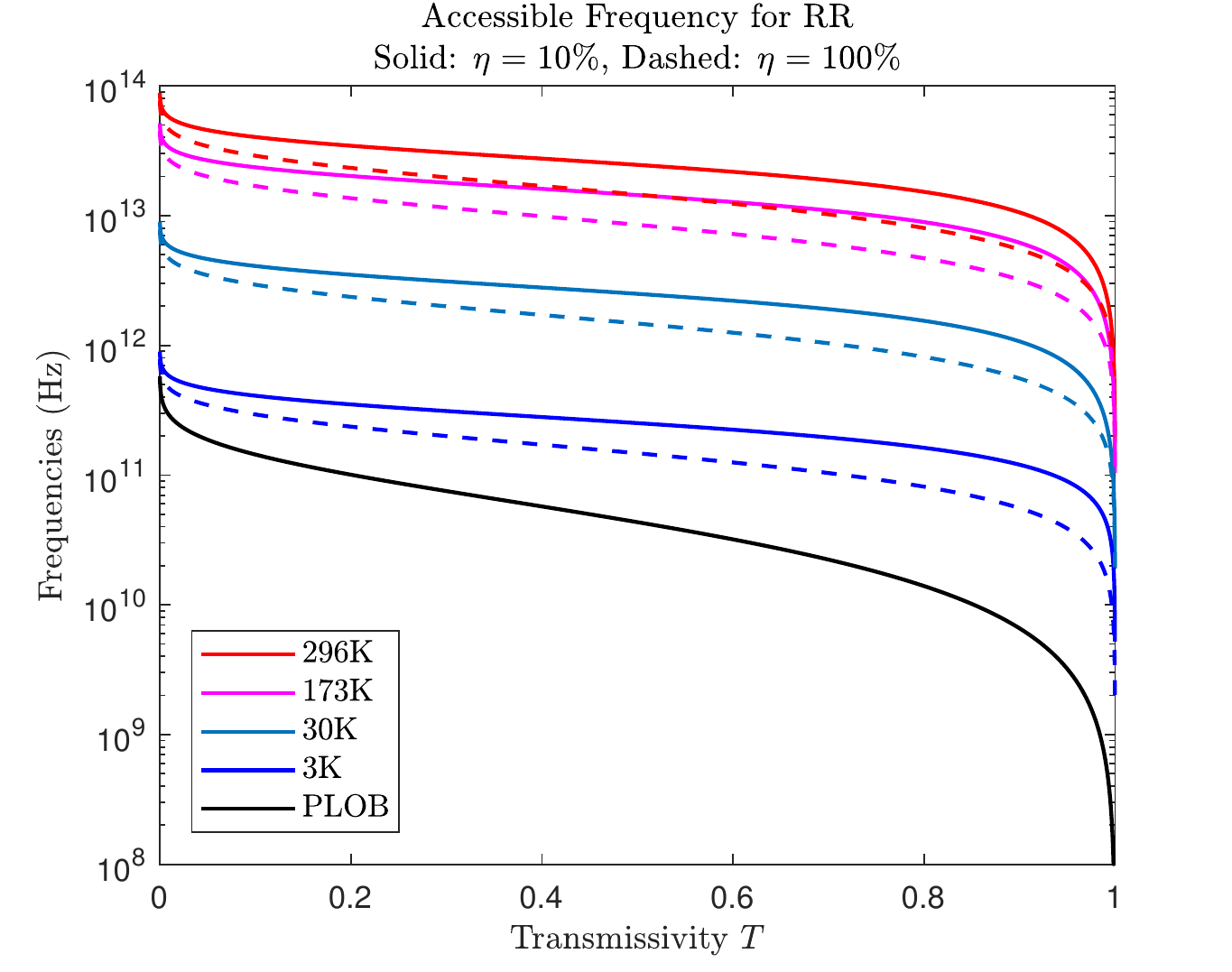}}
		\caption{The accessible frequency against transmissivity for the RR-based inter-satellite QKD protocol at different temperatures. The solid lines refer to the realistic detection efficiency ($\eta=10\%$), and the dashed lines refer to the ideal detection efficiency ($\eta=100\%$). The lower bound on the accessible frequency is calculated at 30K for comparison.}\label{RR_Access_Freq_Fig}
	\end{center}
\vspace{-0.7cm}
\end{figure}

In Fig.~\ref{RR_Access_Freq_Fig} we plot the accessible frequency against channel transmissivity $T$ at the four typical temperatures introduced in Section~\ref{SysModel}. To focus on the inter-satellite QKD scenario, we are mostly interested in 173K and 30K and the results shown are for these temperatures. Also, the realistic (10\%) and an ideal (100\%) detection efficiency are both considered for comparison.

Fig.~\ref{RR_Access_Freq_Fig} shows that the accessible frequencies at our  temperatures of interest are within the THz regime for all possible channel transmissivities. We can see that a lower temperature reduces the accessible frequency, and that a good channel with larger transmissivity can also reduce the accessible frequency. Specifically, when a temperature of 30K is achieved, the maximum accessible frequency for a very bad channel (with $T \sim 0$) is slightly lower than 10THz.  We also notice that a higher detection efficiency can also reduce the accessible frequency.The observations in Fig.~\ref{RR_Access_Freq_Fig} again support the feasibility of inter-satellite THz QKD.

\subsection{Satellite-to-ground QKD}
So far, our discussions have been confined to inter-satellite quantum communications. However, a useful downlink channel to Earth, from the satellite constellation, would provide for enhanced usage  of THz satellite-based QKD.
In such a downlink channel, THz signals will unavoidably traverse the atmosphere, where unfavorable atmospheric turbulence, absorption, and other negative effects are introduced. The higher temperatures at the receiving detector also induces higher noise fluctuations compared to the inter-satellite scenario.

Here we consider a simplified satellite-to-ground QKD model, where the RR-based QKD is carried out between a LEO-based micro-satellite and a ground station which is at a  distance of 500km away. For simplicity, we assume the ground station is located in a dry atmospheric environment such that absorptions can be ignored, and that the satellite-to-ground channel is a diffraction-only channel.
In order to shed light on the feasibility of satellite-to-ground THz QKD, here we try to answer the question: \emph{Under what conditions can a purely-diffracting beam over 500km produce non-zero QKD rate?}

Although the satellite can be cooled down to 30K, the detector on Earth is assumed to be at room temperature (296K), and therefore the total noise of the whole system is dominated by the detection noise at Bob's receiver. For the worst case consideration, we set the variance of Alice's preparation noise equal to the variance of Bob's detection noise. Also, in order to perform the entanglement cloner attack, Eve will match the variance of her TMSV state with the variance of the preparation noise. This translates into setting $V_0=V_s=V_E$ for $T_e = 296$K.

\begin{figure}
	\begin{center}
	\includegraphics[width=3.3 in, height=2.5 in]{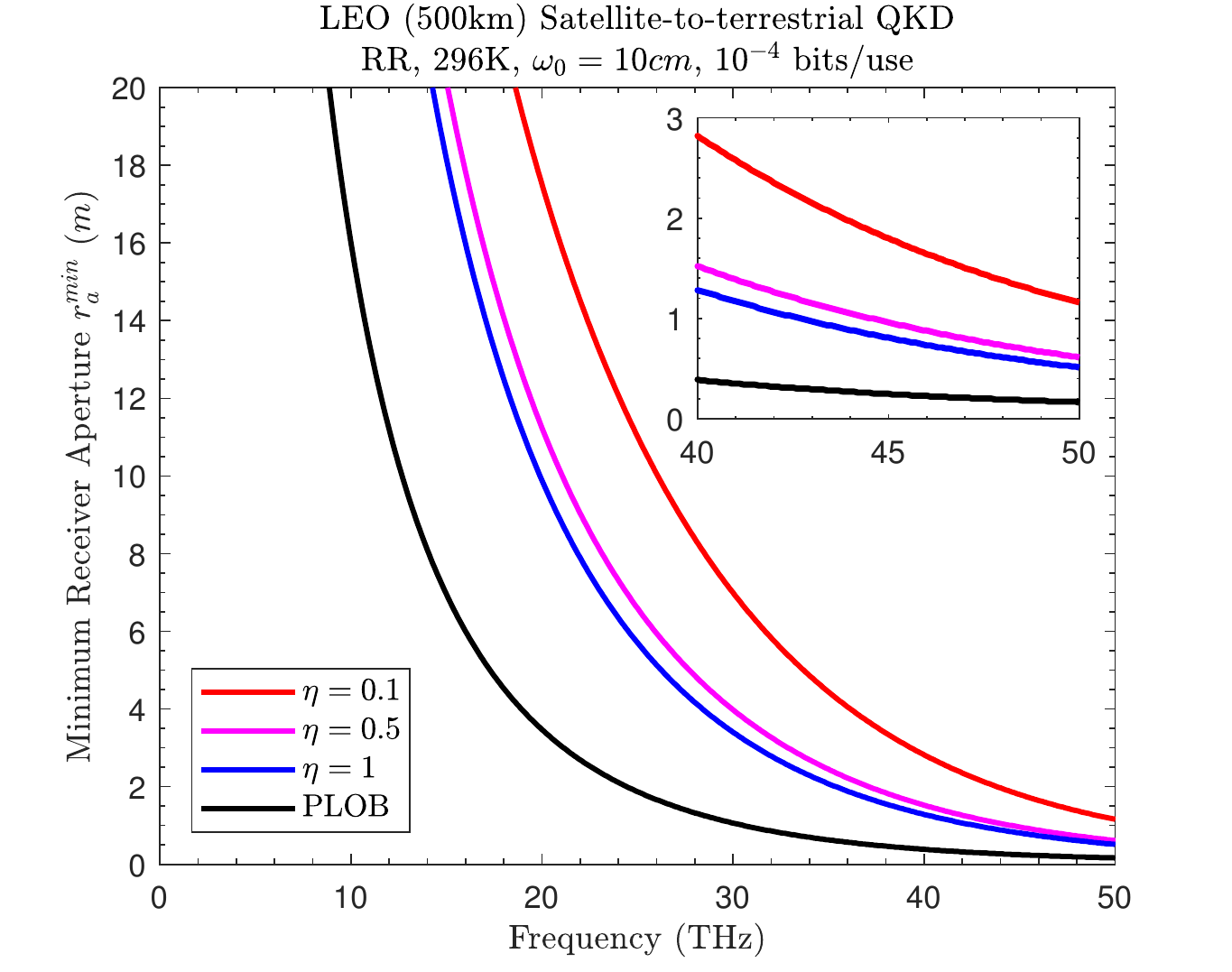}
	\caption{Minimum receiver aperture radius versus QKD frequencies. Multiple detection efficiencies are considered. The PLOB bound with $\eta=1$ is considered for comparison. Here we set $\omega_0=10cm$.}
	\label{fig:satellite-to-groundqkd}
	\end{center}
	\vspace{-0.7cm}
\end{figure}

 We consider the lowest reasonable QKD rate to be  $10^{-4}$ bits/use, and we define $r_a^{min}$ as the minimum receiver aperture  radius required to achieve this minimal QKD rate. $r_a^{min}$ can be found numerically. In Fig.~\ref{fig:satellite-to-groundqkd}, we set $\omega_0$ to 10cm, and plot $r_a^{min}$ against QKD frequencies. Multiple detection efficiencies, namely $\eta=0.1$, $\eta=0.5$, and $\eta=1$, are considered for comparison.

From Fig.~\ref{fig:satellite-to-groundqkd} we can see that a higher QKD frequency can reduce $r_a^{min}$, and so can a higher detection efficiency.
Although $r_a^{min}$ starts to become unrealistic ($>20m$) below 15THz, it is worth noticing that the resulting $r_a^{min}$ for frequencies above 30THz are practical even if the detection efficiency is very poor.

Besides the room temperature scenario, we note that the detector on Earth can be practically cooled to liquid Nitrogen temperatures (77K) especially for ground stations in low humidity locations. According to our simulation, cooling to such a temperature could reduce $r^{a}_{min}$ at 30THz to around 1m and push the feasible QKD frequency down to tens of THz.

\subsection{Quantum radar at THz frequencies}

Quantum Radar (QR) - also referred to as  Quantum Illumination - describes how entangled quantum states can lead to enhanced detection of a remote object via reflection \cite{lloyd, 6dB,microwave_QI}. Deployed in  normal radar operating environments - ambient temperatures at microwave frequencies - QR is anticipated to lead to a 6dB advantage (see below) relative to  coherent-state illumination. We close this work by discussing the nature of the QR advantage operating in the THz regime. We first quantify the QR advantage.

Assume a binary detection where the presence or absence of an object needs to be determined by using a quantum state ${\rho}$. ${\rho}$ is a coherent state for coherent-state illumination, and is a TMSV state for QR. Denote `0' and `1' as  events where the object is present and absent, respectively, and $\mathcal{H}_0$ and $\mathcal{H}_1$ as the two corresponding hypothesis (we assume the \emph{a-priori} probabilities of those two events are equal). The error probability can be given by $\mathrm{Pr}(e)=\frac{1}{2}\mathrm{Pr}(0|\mathcal{H}_1)+\frac{1}{2}\mathrm{Pr}(1|\mathcal{H}_0)$, which is upper-bounded by the quantum Chernoff bound. This upper bound is given by  $\operatorname{Pr}(e) \leq \frac{1}{2}e^{-M\mathcal{E}}\equiv\frac{1}{2}\left\{\min_{0\leq s \leq 1 }\operatorname{tr} \left[ \left( {\rho}^{(0)} \right)^{s} \left( {\rho}^{(1)} \right) ^ {(1-s)} \right] \right\}^{M}$, where ${\rho}^{(0)}$ and ${\rho}^{(1)}$ refer to the quantum states utilized for detection under $\mathcal{H}_0$ and $\mathcal{H}_1$, respectively, $M$ is the number of detection shots, and $\mathcal{E}$ is the error-probability exponent which is used as a metric for the detection performance. The QR advantage in detection performance can be given by $\mathcal{E}_{q}/\mathcal{E}_{c}$, where $\mathcal{E}_{q}$ stands for the error-probability exponent of QR, and $\mathcal{E}_{c}$ and stands for the error-probability exponent of coherent-state illumination. The quantum Chernoff bound is exponentially-tight (viz., $-\ln[\operatorname{Pr}(e)]/M \rightarrow \mathcal{E}$) as $M \rightarrow \infty$. In the results to follow we calculate the QR advantage   in the limit of large $M$.

As noted above,  QR  has a 6dB advantage relative to  coherent-state illumination in a low-reflectivity, low-energy,  high-noise regime, typical of microwave radar deployments at ambient temperatures.
In similar circumstances our calculations show that QR can achieve a 5dB gain in detection performance at 1THz and 3.3dB gain at 10THz. This opens up the possibility of the adoption of QR directly in THz satellite networks without the overhead of  complex microwave-optical interfaces usually envisaged for this technology.

\section{Conclusions}

In this paper, we investigated the feasibility of LEO-based inter-satellite quantum communications using micro-satellites at THz frequencies.

The work presented in this paper opens up the possibility for simpler integration of global quantum and wireless networks. We consider such integration as an enabling step for future global-scale quantum communications.

Despite the  broad prospects for THz quantum communications,  THz-based components (such as the entanglement-generating sources, modulators, and detectors), are less-developed compared to their optical counterparts. One avenue being explored by  researchers is the use of
 THz-optical interfaces as a means of transferring optical quantum information into THz quantum information. However, as mentioned earlier this does induce additional complexity into the overall communication system. `Laser-type' systems embedding quantum information directly into a THz carrier are also currently being explored by  researchers, and these efforts perhaps offer a better quantum-THz integration path.


\end{document}